\documentclass[12pt]{article}
\usepackage{amsmath,amssymb,bm,graphicx}
\usepackage{color} 
\usepackage{float}
\usepackage[lofdepth,lotdepth]{subfig}

\numberwithin{equation}{section}
\usepackage{hyperref}
\usepackage{cite}

\begin{document}


\begin{center}
{\Large{\bf Amorphous solids as fuzzy crystals:\\ A Debye-like theory of low-temperature specific heat}}
\end{center}\vskip .5 truecm
\begin{center}
{\bf \large{  T. Cardoso e Bufalo$^*$, R. Bufalo$^*$ and A. Tureanu$^\dagger$}}

\vspace*{0.4cm} 
{\it {$^*$Departamento de F\'isica, Universidade Federal de Lavras, \\
Caixa Postal 3037, 37200-900, Lavras-MG, Brazil\\
$^\dagger$Department of Physics, University of Helsinki, P.O.Box 64, 
\\FIN-00014 Helsinki,
Finland
}}
\end{center}

\begin{abstract}
We construct a quantum mechanical model of perfectly isotropic amorphous solids as fuzzy crystals and
establish an analytical theory of vibrations for glasses at low temperature. Our theoretical framework relies on the basic principle that the disorder in a glass
is similar to the disorder in a classical fluid, while the latter is mathematically encoded by noncommutative coordinates in the Lagrange formulation of fluid mechanics. 
We find that the density of states in the acoustic branches flattens significantly, leading
naturally to a boson peak in the specific heat as a manifestation of
a van Hove singularity. The model is valid in the same range as Debye's theory, namely up to circa 10\% of the Debye temperature. Within this range, we find an excellent agreement between the theoretical predictions and the experimental data for two typical glasses, a-GeO$_2$ and Ba$_{8}$Ga$_{16}$Sn$_{30}$-clathrate.

Our model supports the conceptual understanding of the nature of the boson peak in glasses as a manifestation of the liquid-like disorder. At the same time, it provides a novel, mathematically simple framework for a unitary treatment of thermodynamical and transport properties of amorphous materials and a solid background for inserting further elements of complexity specific to particular glasses.

\end{abstract}

\section{Introduction}


The theoretical description of the anomalous behavior of amorphous solids  \cite{zeller-pohl-1971} is a recurrent theme in the literature \cite{livro,berthier,Elliot_book}.
A universal anomaly in the frequency spectra of atomic vibrations known as the ``boson peak'' appears at low energies, being usually attributed to an excess in the vibrational density of states (VDOS) compared to the one predicted by the standard Debye theory for crystals ($g_{crystal}(\omega)\sim\omega^2$). It manifests itself in measurements of specific heat and thermal conductivity, as well as scattering of electromagnetic radiation and neutrons. 
The low-temperature ($T < 1$ K) thermal behaviour of glasses is customarily explained
within the established framework of the standard two-level tunneling systems \cite{Phillips,Anderson}, but alternative scenarios have
also been proposed \cite{Leggett}. The anomalous low-temperature properties of glasses  provide indirect evidence regarding the influence of the structural differences on the dynamics governing the vibrations of molecules in the glassy state.
Nevertheless, the physical origin of two most outstanding anomalous behaviors displayed by glasses at intermediate temperatures ($1-40$ K), i.e., the excess of heat capacity and the boson peak, is still under debate.

In this paper, we present a novel theoretical framework for the description of low-temperature vibrations in amorphous solids, which are modelled as "fuzzy crystals". Starting from the general observation that the microscopic structure of a glass is essentially the same as that of a fluid, we use the analytical mechanics tool for describing fluid disorder as a coordinate-reparametrization symmetry, leading to nonvanishing Poisson brackets between the coordinates \cite{Susskind-Bahcall, Susskind, Jackiw1}. In intuitive terms, this mathematical structure induces a fuzziness of the coordinate space, which blurs the points and produces an uncertainty in the localization -- in short, fluid-specific disorder. When superimposed on a crystalline lattice, this results into a fuzzy crystal.
The analytical glass model constructed in this way naturally contains the boson peak as a manifestation of an extended van Hove singularity.
The key concept is the quantization of the configuration space, as an abstract realization of the disorder in the glass.

The noncommutativity of coordinates has a long history in theoretical physics, appearing in the Hamiltonian description of systems with constraints, when passing from Poisson to Dirac brackets \cite{Dirac-book}. The first examples appeared in condensed matter physics, namely Peierls' model of a charged particle in a strong magnetic field \cite{Peierls}, which is connected also to the behaviour of charged particles in the lowest Landau level. A similar set-up with an antisymmetric B-field, in string theory, gives quantum field theory on a space-time with noncommuting coordinates as a low-energy effective limit \cite{SW}. 
During the last two decades, the noncommutativity of space(-time) has been an active field of research in high-energy physics and string theory.
On the other hand, the noncommutativity of coordinates as a manifestation of the coordinate-reparametrization symmetry in a Lagrange fluid has been explored in other condensed matter systems, for example in the description of the quantum Hall fluid in terms of a noncommutative Chern-Simons quantum field theory\cite{Susskind,Hellerman,Lapa:2018ubk}.
Our proposal for modeling amorphous materials and the noncommutative formulation of the quantum Hall fluid have the same underlying framework.

In Sec.\ref{sec.2}  we review the Lagrangian description of fluids, and show how, by imposing invariance under reparametrization of the coordinates, one arrives at a set of noncommuting coordinates \cite{Susskind-Bahcall, Susskind, Jackiw1}.
The quantum algebra of these new variables introduces an additional {\it uncertainty relation}, which manifests itself as a ``fuzziness'' of space points, or an intrinsic  ``disorder'', that is essential in our glass model developed in Sec. \ref{sec.3}.
There we establish the main aspects of the description of glasses in terms of the noncommutative fluid picture. In Sec. \ref{debye} we establish the Debye-like theory for specific heat of glasses and derive the density of states.
In Sec. \ref{sec.4} we compute and discuss the reduced low-temperature specific heat, showing that the fuzzy crystal model yields the expected profile for glasses: excess of heat capacity and the boson peak.
We present our final remarks and prospects in Sec. \ref{sec.5}.

\section{Fluids in Lagrangian description}
\label{sec.2}
A fluid can be described in the Lagrangian picture (see, e.g., \cite{Mechanics}), in which one follows the motion of each individual fluid particle. Assuming a two-dimensional fluid composed of
$N$ identical particles, in which the coordinate of particle $i$ at the time
$t$ will be $\mathbf{X}_{i}(t)$, satisfying the Newton equation
\begin{equation}
\ddot{\mathbf{X}}_{i}(t)=\frac{1}{m}\mathbf{F}(\mathbf{X}_{1},\ldots ,%
\mathbf{X}_{N}).  \label{1}
\end{equation}%
Thus the Lagrangian of the system will be
\begin{equation}
L(t)=\sum_{i=1}^{N}\frac{1}{2}m\dot{\mathbf{X}}_{i}(t)-V(\mathbf{X}%
_{1},\ldots ,\mathbf{X}_{N}),  \label{2}
\end{equation}%
such that in the passage from the discrete description to a continuous one,
the label $i$ takes continuous values denoted by $\mathbf{x}$. Therefore the
Lagrangian coordinate $\mathbf{X}_{i}(t)$ will become the vector field $%
\mathbf{X}(\mathbf{x},t)$, keeping the customary interpretation of the
continuous label $\mathbf{x}$ as the value of the coordinate $\mathbf{X}$ at
the initial moment $t=0$, i.e.
\begin{equation}
\mathbf{X}(\mathbf{x},0)=\mathbf{x}.  \label{3}
\end{equation}%
Actually, the $\mathbf{x}$-coordinates represent a reference configuration
of the fluid, not necessarily attained during its motion. Since the $\mathbf{%
x}$s label the particles of fluid, they move with them, therefore the system
of $\mathbf{x}$ coordinates is a \textit{comoving} system. These coordinates
are chosen so that the number of particles per unit is constant and equal to
$\rho _{0}$. Denoting by $\rho $ the density in the (real) space of $\mathbf{%
X}$ coordinates, it follows that
\begin{equation}
\rho =\rho _{0}\left\vert \frac{\partial \mathbf{x}}{\partial \mathbf{X}}%
\right\vert .  \label{4}
\end{equation}%
The Lagrangian in the continuous description becomes
\begin{equation}
L(t)=\int d\mathbf{x}\left( \frac{1}{2}m\dot{\mathbf{X}}(\mathbf{x},t)-%
\mathcal{V}(\mathbf{X}(\mathbf{x},t))\right) .  \label{5}
\end{equation}

Assuming that the fluid is a continuum, it has been shown \cite{Susskind-Bahcall, Susskind, Jackiw1} that the fluid dynamics encoded in the Lagrangian of the system \eqref{5} is invariant under a reparametrization of the coordinates, which preserves the surface density of particles $\sigma_{0}$ in any plane of the fluid (in the discrete description, this is equivalent to invariance under renaming the arbitrary particle labels). Such
transformations are volume-preserving diffeomorphisms:
\begin{equation}\label{6}
\mathbf{x}\rightarrow \mathbf{x}^{\prime }=\mathbf{x}+\delta \mathbf{x},\ \
\ \ \delta \mathbf{x}=\mathbf{f}(\mathbf{x}).
\end{equation}
The area-preservation condition $\left\vert \frac{\partial \mathbf{x}%
^{\prime }}{\partial \mathbf{x}}\right\vert =1$ leads to
\begin{equation}
\nabla \cdot \mathbf{f}(\mathbf{x})=0,  \label{7}
\end{equation}%
i.e. the infinitesimal vector function $\mathbf{f}$ has to be transverse. For simplicity, we consider a two-dimensional space. Then the latter condition can be written in terms of a scalar function $f$ as
\begin{equation}\label{8}
f_{i}(\mathbf{x})=\epsilon _{ij}\frac{\partial f}{\partial x_{j}}.
\end{equation} 
It has been proven (see also \cite{Jackiw2, Polyn}) that in this approach, the reparametrization symmetry can be re-cast in the form of noncommuting coordinates, namely by introducing additional Poisson brackets
\begin{equation}
\{x_{i},x_{j}\}=\theta _{ij},  \label{9}
\end{equation}%
with an arbitrary set of constants $\theta_{ij} $. The isotropy of the fluid is ensured by the arbitrariness of the elements of the matrix $\theta_{ij} $. Rescaling $f$ by $\theta^{-1}$, we can re-write $\partial x_i$ as
\begin{equation}\label{10}
\delta x_i=\theta_{ij}\partial_j f=\{x_i,f\}.
\end{equation}
The Poisson brackets \eqref{9} are invariant under the transformations \eqref{10}, the elements $\theta_{ij}$ remaining invariant under the reparametrization of coordinates \cite{Jackiw1}. The proof is based
on the fact that the Lie derivative of the tensor $\theta _{ij}$ with
respect to the vector field $\mathbf{f}$ vanishes. Under the area-preserving
diffeomorphisms \eqref{10}, the Lagrangian coordinates $\mathbf{X}$
transform as
\begin{equation}
\delta X_{i}=\partial _{j}X_{i}\delta x_{j}=\theta _{jk}\partial
_{j}X_{i}\partial _{k}f=\{X_{i},f\}.  \label{11}
\end{equation}
The generalization to three- and higher-dimensional spaces is straightforward.

In the quantum treatment of the fluid, the Poisson bracket of coordinates
\eqref{9} becomes the commutator
\begin{equation}
\lbrack \widehat x_{i},\widehat x_{j}]=i\theta _{ij}, \quad i,j=x,y,z,
\label{12}
\end{equation}%
(with an antisymmetric matrix $\theta_{ij}$), which has to be added to the usual canonical commutation relations
\begin{equation}
\lbrack \widehat x_{i},\widehat p_{j}]=i\hbar \delta _{ij},\quad [\widehat p_{i},\widehat p_{j}]=0.  \label{13}
\end{equation}
The theory based on the commutators \eqref{12} and \eqref{13} is customarily called noncommutative quantum mechanics.

To the matrix $\theta_{ij}$ we can associate a vector $\vec \theta$, whose components are given by $\theta_i=\epsilon_{ijk}\theta_{jk}$. We note that the volume-preserving diffeomorphisms leave invariant
simultaneously $\vec \theta $ and $\sigma _{0}$. We expect a relation between these two invariants, which can be derived by analogy with
usual quantum mechanics. In the latter case, the commutation relation
$[\widehat x,\widehat p]=i\hbar $ leads to a quantization of the
two-dimensional phase space in cells of area $2\pi \hbar $. In the case of noncommuting
coordinates, the commutator \eqref{12} leads to a {\it quantization}
 {\it of the reference configuration space in cells of area} $2\pi \theta $. We attribute to this basic quantum of area the meaning of surface ``occupied'' by a single particle \cite{Susskind}. But  the inverse density of particles has the same
significance, leading to the relation
\begin{equation}
\theta =\frac{1}{2\pi \sigma _{0}}.  \label{14}
\end{equation}
Incidentally, this approach has been applied to the quantum Hall fluid, providing an alternative description to Laughlin's theory by a noncommutative Chern--Simons quantum field theory \cite{Susskind}.

Let us dwell for a while on the physical significance of the commutator \eqref{12}: its presence in the quantum algebra introduces an additional {\it uncertainty relation}, which manifests itself as a ``blurriness'' of space points. Precise localization, even theoretical, of the particles becomes impossible.
This suggests an intrinsic  ``disorder''. As we shall see further, the interactions are profoundly affected by the coupling of vibrational modes specific to noncommutative quantum mechanics.

The interactions in noncommutative quantum mechanics are described by the Schr\"odinger equation for $N$ degrees of freedom, with the Hamiltonian
\begin{equation}\label{Hamilt_gen}
\widehat H \equiv \sum_i^N \frac{\widehat p_i \widehat p_i}{2M}+ V\left( \widehat x_1,\ldots, \widehat x_N\right),
\end{equation}
where the canonical coordinates $ (\widehat{x}_i,\widehat{p}_j)$, $i,j=1,\ldots,N$ satisfy the extended Heisenberg algebra \eqref{12}-\eqref{13}. The mathematical manipulations are considerably simplified by observing (see, e.g., \cite{CDP_98, Bigatti_Susskind}) that the shifted coordinates
\begin{equation}\label{Bopp}
\widehat X_{i} = \widehat x_i + \frac{1}{2\hbar} \theta_{ij}\widehat p_{j},\quad
\widehat{P}_{i}= \widehat p_{i}
\end{equation}
satisfy the usual Heisenberg algebra $[\widehat X_i,\widehat P_j]=i\hbar\delta_{ij}$, $[\widehat X_i,\widehat X_j]=0$, $[\widehat P_i,\widehat P_j]=0$ (summation over the repeated index is assumed in \eqref{Bopp}).
This allows a physical interpretation of the $\theta$-term in \eqref{Bopp} as a quantum shift operator  \cite{Anca}, namely, a translation in space by $\frac{1}{2\hbar} \theta_{ij}\widehat p_{j}$, while $X_{i}$ is the classical geometrical coordinate.
This technique has been used for deriving noncommutative space corrections to various quantum mechanical phenomena, for example, the Lamb shift \cite{Tureanu}.
If the potential energy in \eqref{Hamilt_gen} is of the harmonic oscillator type, then, when the shift \eqref{Bopp} is applied to $H \left(\widehat{ x},\widehat{ p}\right)$, one obtains a Hamiltonian $H (\widehat X,\widehat P)$ consisting of a sum of usual quantum mechanical harmonic oscillators plus an additional interaction term proportional to the noncommutativity parameter.

\section{Fuzzy crystal model}
\label{sec.3}

The glass model we propose essentially relies on the noncommutative fluid picture.
Glasses are not ordinary liquids due to their {\it rigidity}, nor regular solids, due to their {\it disorder}.
The rigidity is introduced by means of a simple cubic lattice (see also \cite{27}).
The disorder reminiscent of the fluid which was quenched into the glass will be implemented through the noncommutative quantum algebra \eqref{12}-\eqref{13}.

The model is formulated on the basis of several natural assumptions:

\begin{itemize}
\item The glass is composed of a {\it simple cubic lattice} of motifs of mass $M$, with the unit cell vector $a$;

\item The interactions in the disordered lattice are described by a noncommutative harmonic oscillator potential function and we consider harmonic interactions only between the first neighbors of the atoms in a lattice;

\item The vector $\vec\theta$ has equal projections denoted by $\theta$ on the coordinate axes, i.e. on the directions of the edges of the unit cells.
\begin{equation}
\theta _{12}=\theta _{23}=\theta _{31}=\theta .  \label{18}
\end{equation}%
With this assumption all the directions in the cubic lattice are treated in
the same way, there is no preferred direction. 
\end{itemize}

Assuming harmonic interactions among the nearest neighbors of a simple cubic
lattice composed of neutral identical atoms, the usual (commutative) basic
model for normal modes of vibration of a simple cubic lattice can be determined
from the three-dimensional Hamiltonian $H = \frac{\hat p \cdot \hat p}{2m}+
V\left( \hat u\right)$, where $\hat u$ are the generalized displacements	 that will be specified further. The potential energy
in the harmonic approximation is $V\left(\hat u\right) = \frac{1}{2} k \hat
u \cdot \hat u$.

Let us analyze the atom indexed by the integer labels ${lmn}$.
The quantum coordinate operators which define it are 
\begin{eqnarray}
\widehat x_{lmn}&=&{l}\,a+\widehat u^x_{lmn},\cr
\widehat y_{lmn}&=&{m}\,a+\widehat u^y_{lmn},\cr
\widehat z_{lmn}&=&{n}\,a+\widehat u^z_{lmn},
\end{eqnarray} 
where $\widehat u^i$, with $i= x,y,z$, denote generically the displacement operators from the lattice site in the corresponding direction.
The Hamiltonian governing the time evolution of the atom $({lmn})$ is then
\begin{align} \label{ham_1}
\widehat H_{lmn} & =\sum_{i=x,y,z}\frac{1}{2M}\left(\widehat{p}^i_{lmn}\right)^{2} \\
& +\frac{M\omega_{0}^{2}}{2}\left[\left(\widehat u^x_{lmn}-\widehat u^x_{l-1mn}\right)^{2}+\left(\widehat u^x_{l+1mn}-\widehat u^x_{lmn}\right)^{2}\right]\cr
& +\frac{M\omega_{0}^{2}}{2}\left[\left(\widehat u^y_{lmn}-\widehat u^y_{lm-1n}\right)^{2}+\left(\widehat u^y_{lm+1n}-\widehat u^y_{lmn}\right)^{2}\right]\cr
& +\frac{M\omega_{0}^{2}}{2}\left[\left(\widehat u^z_{lmn}-\widehat u^z_{lmn-1}\right)^{2}+\left(\widehat u^z_{lmn+1}-\widehat u^z_{lmn}\right)^{2}\right]\nonumber,
\end{align}
where $\widehat{p}_{lmn}^{i}$ is the momentum canonically conjugated to the displacement $\widehat u^i_{lmn}$ and $\omega_0=\sqrt\frac{k}{M}$ is the usual harmonic oscillator frequency.

\begin{figure}[t!]
\centering
\subfloat[Subfigure 1 list of figures text][Dispersion relations for $R = 0.15$]  
 {\
\includegraphics[width=7cm]{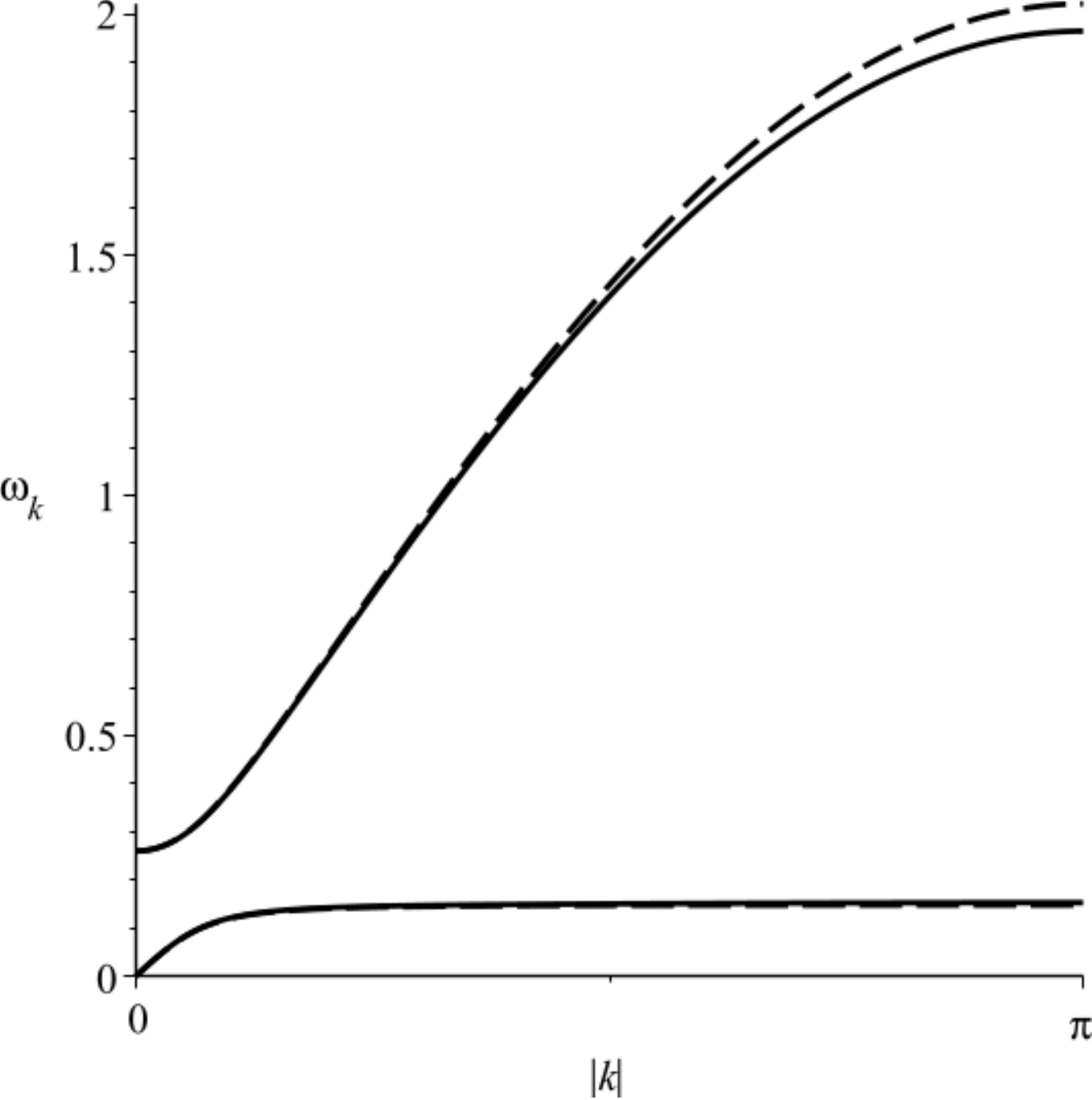}} ~
\subfloat[Subfigure 1 list of figures text][Dispersion relations for $R = 0.45$]  
{\ \includegraphics[width=7cm]{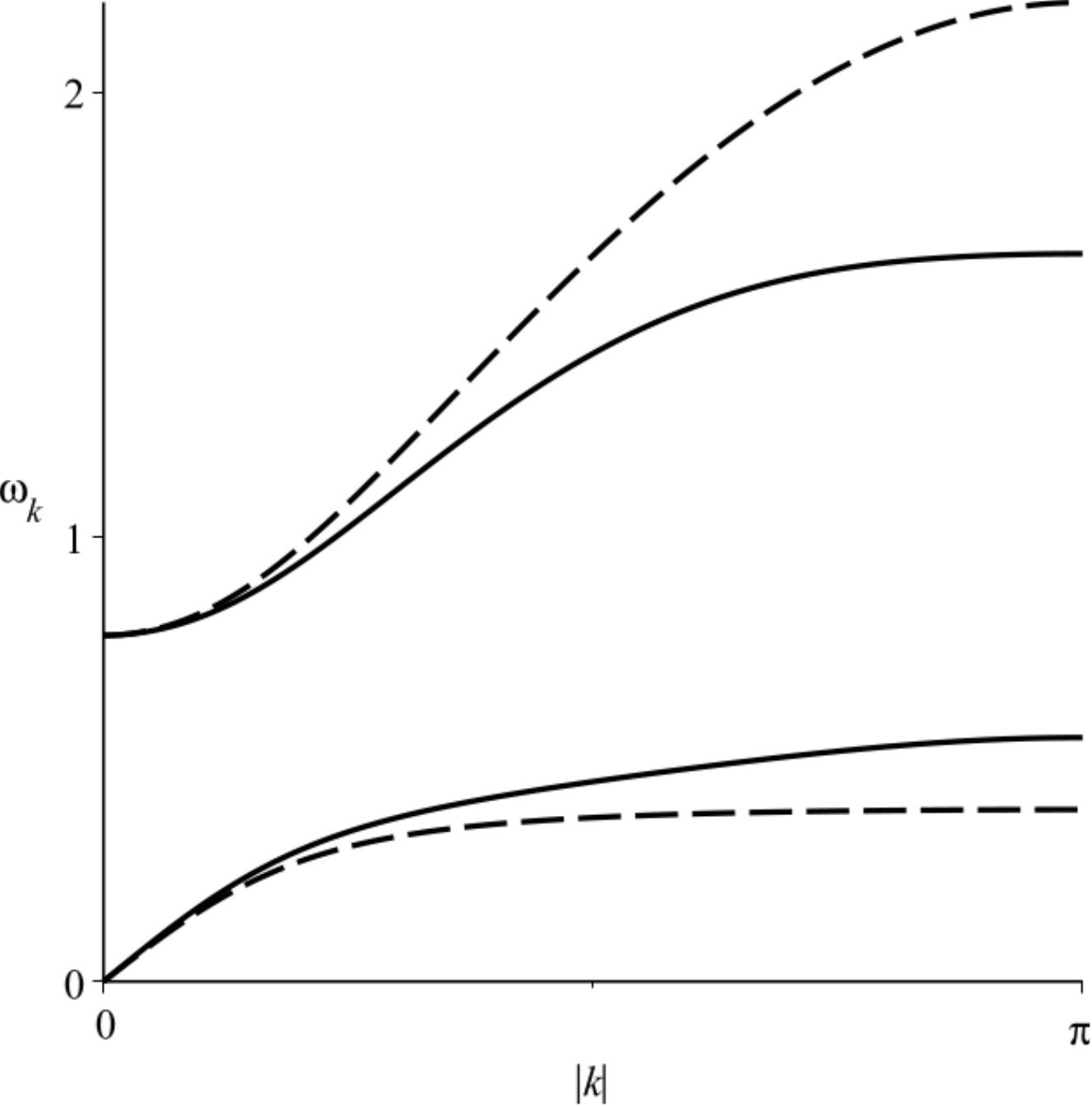}} \newline
\caption{\textcolor{black}{\textbf{Dispersion relations for $R = 0.15$ and $R = 0.45$.}
{Dispersion relations for disordered (dashed lines) and
ordered (solid lines) atoms.
The degenerated acoustic branches are those for which $\omega_k = 0$ as $|k| = 0$, while the optical modes are characterized by $\omega_k \neq 0$ as $|k| = 0$. Note that for very small $R$, the flattening of the dispersion curves in the acoustic branches is almost identical for ordered and disordered atoms.}
}}
\label{completapprox_2}
\end{figure}

In order to make sure that the disorder effects are not doubly-counted, nor washed out, we make an additional assumption about the interactions of the atoms on the lattice.
Namely, we consider an alternation of ordered and disordered atoms.
By {\it ordered atoms}, we mean atoms whose quantum coordinates and momenta satisfy the usual Heisenberg algebra, i.e. whose coordinate operators commute.
By {\it disordered atoms}, we mean atoms whose coordinates and momenta satisfy the noncommutative space algebra \eqref{12}-\eqref{13}.
The arrangement of the lattice is such that one ordered atom is surrounded by disordered first neighbors and vice-versa. 
Due to the presence of elastic couplings, the equations of motion of the so-called ordered atoms will be also influenced by the disorder of their neighbors, such that the lattice as a whole will be disordered. Consequently, all the constituents of the lattice are genuinely disordered. As we shall see further (Fig. \ref{completapprox_2}), for small values of the parameter $\theta$, the difference in the dispersion relations between the two types of atoms is extremely small.

Let us specify the Hamiltonian \eqref{ham_1} for the two types of atoms.
Considering that the atom $lmn$ is a disordered one, which is subject to the effects of the noncommutativity of coordinates, we perform the Bopp shift \eqref{Bopp} by replacing in \eqref{ham_1}
\begin{equation}  \label{eqdis}
\widehat{u}^i_{lmn}=\widehat U^i_{lmn}-\frac{1}{2\hbar}\theta_{ij}\widehat P_{lmn}^j,
\end{equation}
while for all the displacements of the nearest neighbors we have $\widehat u^i_{l+1mn}= \widehat U^i_{l+1mn}$ etc.

On the other hand, if the generic atom $lmn$ is a ordered one, its displacements are $\widehat u^i_{lmn}=\widehat U^i_{lmn}$, while the shifts \eqref{Bopp} have to be performed for the coordinates of the neighbors  in the Hamiltonian \eqref{ham_1}, e.g.
\begin{equation}   \label{eqord}
\widehat{u}^x_{l\pm1mn}=\widehat U^x_{l\pm1mn}-\frac{1}{2\hbar}\theta_{xj}\widehat P_{l\pm1mn}^{j}, \quad  j=y,z.
\end{equation}
In all cases, $\widehat{p}_{lmn}^{i}=\widehat P_{lmn}^{i} $.

The equations of motion for the displacements $\widehat u^i_{lmn}$ follows from the Heisenberg equations, taking into account the canonical commutation relations
\begin{align}  \label{29}
\left[\widehat U_{lmn}^{i},\widehat P_{l^{\prime }m^{\prime }n^{\prime }}^{i'}\right]
=i\hbar\delta^{ii'}\delta_{ll^{\prime }}\delta_{mm^{\prime }}\delta_{nn^{\prime }},\cr
\left[\widehat U_{lmn}^{i},\widehat U_{l^{\prime }m^{\prime }n^{\prime }}^{i'}\right]=\left[\widehat P_{lmn}^{i},\widehat P_{l^{\prime }m^{\prime }n^{\prime }}^{i'}\right]=0.
\end{align}
These considerations are imperative to determine the density of states for the complete system within Born--von K\'{a}rm\'{a}n's procedure \cite{Kittel}, which ultimately allow us to compute the specific heat for the model.
Hence, through this section we shall determine the equations of motion and the corresponding dispersion relations, that allow us to compute the density of states for the glass model.

A last comment is that so far, the model has two sources of anisotropy: the crystal lattice and the vector  $\vec{ \theta}$.
As far as the noncommutativity is concerned, the disorder effect is manifest only in the plane orthogonal to $\vec{\theta}$, and not in the direction of $\vec{ \theta}$.
In order to render the model isotropic, we choose as representative axis one of the coordinates axes (e.g. $Oz$), and determine the dispersion relations for it, $\omega_{k_z}(k_z)$.
With our choice of $\vec{\theta}$ having equal projections on the coordinate axes, we insure that the disorder induced by noncommutativity is similar along $Oz$ and in the plane orthogonal to it. Subsequently, we replicate the $Oz$ axis by rotational symmetry to all the directions of the Cartesian frame, i.e. $\omega_{k_z}(k_z) \rightarrow \omega_{\mathbf{k}}(|\mathbf{k}|)$, by replacing in $\omega_{k_z}(k_z)$ everywhere $k_z$ by $|\mathbf{k}|$.

\section{Debye-like theory for specific heat of glasses}\label{debye}

Our purpose is to calculate the specific heat of glasses at low temperature using the "frozen fluid" quantum mechanical model described in the previous section. We shall follow the technical steps of the Debye theory of crystals, namely find the density of states $g_{glass}(\omega)$ within our model, which contribute to the internal energy by the formula
\begin{equation}
U=\int d\omega \hbar \omega n_{BE}(\omega,T) g_{glass}(\omega),
\end{equation}
where $n_{BE}(\omega,T)$ is the Bose--Einstein distribution. By differentiating this equation with respect to the temperature, we find the specific heat $C$

\subsection{Density of states for disordered atoms}


The equations of motion for the two kinds of atoms are obtained via Heisenberg's equations
\begin{equation}
i\hbar \frac{d\widehat{O}\left( t\right) }{dt}=\left[ \widehat{O}\left(
t\right) ,\widehat H\right] .  \label{28}
\end{equation}%
In order to determine the equations of motion for a disordered atom we shall consider our previous discussion: start by rewriting the Hamiltonian \eqref{ham_1} in terms of a new set of variables, $(u,p)\to(U,P)$, where the displacements for the disordered atoms are obtained from the shift \eqref{eqdis}, while the displacements of the nearest (ordered) neighbors are simply $\widehat u^i_{l+1mn}= \widehat U^i_{l+1mn}$.

Hence, under these considerations, the equations of motion for a disordered atom $(lmn)$ are
\begin{align}
\ddot{\widehat{U}}_{lmn}^{x} &=-\omega _{0}^{2}\left( 2\widehat{U}_{lmn}^{x}-%
\widehat{U}_{l-1mn}^{x}-\widehat{U}_{l+1mn}^{x}\right) \cr
&+\frac{\omega _{0}R}{2
}\left[ \left( 4\dot{\widehat{U}}_{lmn}^{y}-\dot{\widehat{U}}_{lm-1n}^{y}-%
\dot{\widehat{U}}_{lm+1n}^{y}\right) -\left( 4\dot{\widehat{U}}_{lmn}^{z}-%
\dot{\widehat{U}}_{lmn-1}^{z}-\dot{\widehat{U}}_{lmn+1}^{z}\right) \right], \cr
\ddot{\widehat{U}}_{lmn}^{y}&=-\omega _{0}^{2}\left( 2\widehat{U}%
_{lmn}^{y}-\widehat{U}_{lm-1n}^{y}-\widehat{U}_{lm+1n}^{y}\right) \cr
&+\frac{%
\omega _{0}R}{2}\left[ \left( 4\dot{\widehat{U}}_{lmn}^{z}-\dot{\widehat{U}}%
_{lmn-1}^{z}-\dot{\widehat{U}}_{lmn+1}^{z}\right) -\left( 4\dot{\widehat{U}}%
_{lmn}^{x}-\dot{\widehat{U}}_{l-1mn}^{x}-\dot{\widehat{U}}%
_{l+1mn}^{x}\right) \right], \cr
\ddot{\widehat{U}}_{lmn}^{z}&=-\omega
_{0}^{2}\left( 2\widehat{U}_{lmn}^{z}-\widehat{U}_{lmn-1}^{z}-\widehat{U}%
_{lmn+1}^{z}\right) \cr
&+\frac{\omega _{0}R}{2}\left[ \left( 4\dot{\widehat{U}}%
_{lmn}^{x}-\dot{\widehat{U}}_{l-1mn}^{x}-\dot{\widehat{U}}%
_{l+1mn}^{x}\right) - \left( 4\dot{\widehat{U}}%
_{lmn}^{y}-\dot{\widehat{U}}_{lm-1n}^{y}-\dot{\widehat{U}}%
_{lm+1n}^{y}\right)   \right] ,  \label{eom_D}
\end{align}
where we have introduced the notations
\begin{equation}\label{R}
\omega_{\theta}= \frac{\hbar}{M \theta}\quad \mbox{and}\quad R=\frac{\omega_0}{\omega_\theta}.
\end{equation}
The latter is a dimensionless expansion parameter, proportional to $\theta$.
In the computation of the equations of motion \eqref{eom_D} we have retained only terms up to the second order in $R$, as this parameter is expected to be very small. 

Any function in a space formed by a periodic arrangement of atoms must satisfy periodic boundary conditions, the Born--von K\'{a}rm\'{a}n boundary conditions.
It is important to highlight that the periodicity of the lattice is maintained under the shift of coordinates applied to the quantum Hamiltonian, since the disorder effects emerge as a new interaction terms added to the ordinary Hamiltonian of the simple cubic lattice.
Therefore, under these conditions, the Born--von K\'{a}rm\'{a}n boundary conditions can be freely applied in our description.

Hence, within our approach, we consider Born's approach to the vibrations of a lattice, expanding in Fourier series the operators
\begin{equation}\label{Fourier_solution}
\widehat U ^i_{lmn}=\sum_{\mathbf{k}}\hat{\cal U}^{i}_{\mathbf{k}}\exp \left[ i\left( \omega_{\mathbf{k}} t+a(lk_x+mk_y+n k_z) \right)\right],
\end{equation}
using the Born--von K\'{a}rm\'{a}n boundary conditions (see, e.g., \cite{Landau,Kittel,Kantorovich}).
In the above expression, $k_i$ represent the wave vector projections and $\omega_{\mathbf{k}}$ the mode vibration frequencies.
We obtain the dispersion relations by inserting the expansion \eqref{Fourier_solution} into the equations of motion.

Replacing the Ansatz \eqref{Fourier_solution} into the equations of motion \eqref{eom_D},
we find the saecular equation for the disordered atoms:
\begin{equation}
\left\vert
\begin{array}{ccc}
\omega_{\mathbf{k}} ^{2}-4\omega _{0}^{2}\sin ^{2}\left( \frac{ak_{x}}{2}%
\right) & i\omega_{\mathbf{k}} \omega _{0}R\left[ 1+2\sin ^{2}\left( \frac{%
ak_{y}}{2}\right) \right] & -i\omega_{\mathbf{k}} \omega _{0}R\left[ 1+2\sin
^{2}\left( \frac{ak_{z}}{2}\right) \right] \\
-i\omega_{\mathbf{k}} \omega _{0}R\left[ 1+2\sin ^{2}\left( \frac{ak_{x}}{2}%
\right) \right] & \omega_{\mathbf{k}} ^{2}-4\omega _{0}^{2}\sin ^{2}\left(
\frac{ak_{y}}{2}\right) & i\omega_{\mathbf{k}} \omega _{0}R\left[ 1+2\sin
^{2}\left( \frac{ak_{z}}{2}\right) \right] \\
i\omega_{\mathbf{k}} \omega _{0}R\left( 1+2\sin ^{2}\left( \frac{ak_{x}}{2}%
\right) \right) & -i\omega_{\mathbf{k}} \omega _{0}R\left[ 1+2\sin
^{2}\left( \frac{ak_{y}}{2}\right) \right] & \omega_{\mathbf{k}}
^{2}-4\omega _{0}^{2}\sin ^{2}\left( \frac{ak_{z}}{2}\right)%
\end{array}%
\right\vert =0 .  \label{saec_eq_disordered}
\end{equation}
As previously explained, the only physically relevant directions for our model are the directions of coordinates.
The isotropization of the model is performed by taking the dispersion relations in the direction $Oz$ and replicating it by rotational symmetry to all the directions of the Cartesian system, namely by the replacement $k_z\to |{\mathbf{k}}|$.

Upon isotropization, the solutions of equation \eqref{saec_eq_disordered} (for $k_x=k_y\equiv 0$ and $k_z\to {\mathbf{k}}$), are:
\begin{align} 
\omega _{{\mathbf{k}}\pm }^{\mathrm{D}} =&\frac{\omega _{0}}{2}\bigg[
6R^{2}+8\left( 1+R^{2}\right) \sin ^{2}\left( \frac{a|{\mathbf{k}}|}{2}%
\right)   \cr
& \pm 2\bigg[ -7R^{4}-32R^{2}-16  + 8\left( 7R^{4}+9R^{2}+4\right) \sin ^{2}\left( \frac{a|{%
\mathbf{k}}|}{2}\right) \cr
& + 16\left( 1+R^{2}\right)^2 \left( 1-\sin
^{2}\left( \frac{a|{\mathbf{k}}|}{2}\right)  \right)^2 \bigg] ^{\frac{1}{2}}\bigg] ^{\frac{1}{2}},  \label{omega_D}
\end{align}
where the minus (plus) sign is related to the degenerated acoustic (optical) branch.

In the small angle approximation (i.e. $\sin (ak_{i})\approx ak_{i}$), the saecular equation \eqref{saec_eq_disordered} becomes:
\begin{equation}
\omega_{\mathbf{k}} ^{4}+\omega _{0}^{2}\left[ \omega _{0}^{2}R^{2}-\omega_{%
\mathbf{k}} ^{2}\left( 1+R^{2}\right) \right] a^{2}|\mathbf{k}|^2 -3\omega_{%
\mathbf{k}} ^{2}\omega _{0}^{2}R^{2}=0,  \label{disordered_se}
\end{equation}
with the solution
\begin{align}
\omega _{\mathbf{k}\pm}^{\mathrm{D-approx}}= &\frac{\omega _{0}}{2}\bigg[
6R^{2}+2\left( 1+R^{2}\right) a^{2}|{\mathbf{k}}|^{2} \cr
&\pm 2\sqrt{%
9R^{4}+2R^{2}\left( 3R^{2}+1\right) a^{2}|{\mathbf{k}}|^{2}+\left(
R^{2}+1\right) ^{2}a^{4}|{\mathbf{k}}|^{4}}\bigg]   \label{omega_D_app}
\end{align}
preserving the relation of the minus (plus) sign for the acoustic (optical)
branch.

\begin{figure}[t!]
\centering
\subfloat[Subfigure 1 list of figures text][Zoom at the acoustic branch of disordered atoms.]  
 {\
\includegraphics[width=7cm]{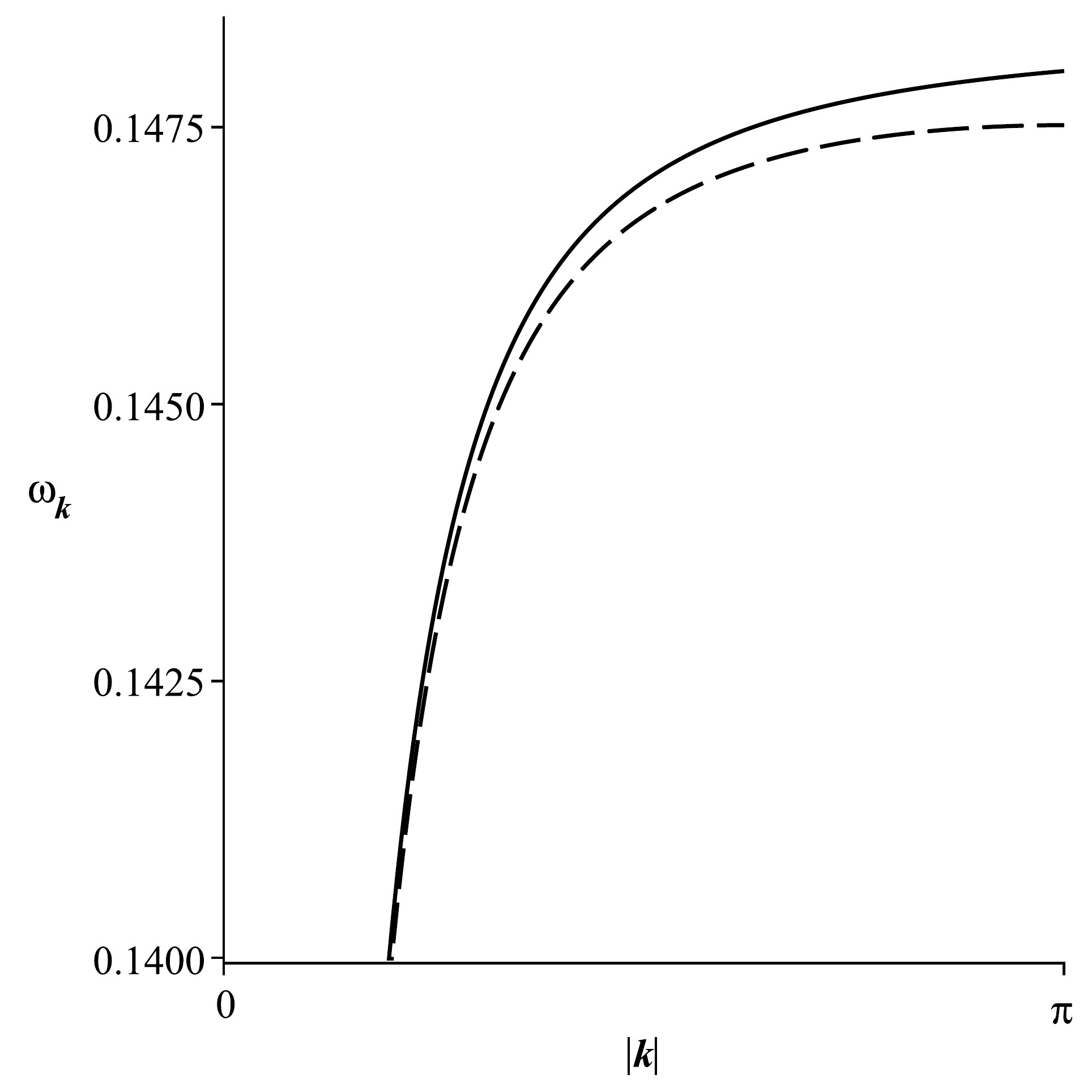}} ~
\subfloat[Subfigure 1 list of figures text][Zoom at the acoustic branch of ordered atoms.] 
{\ \includegraphics[width=7cm]{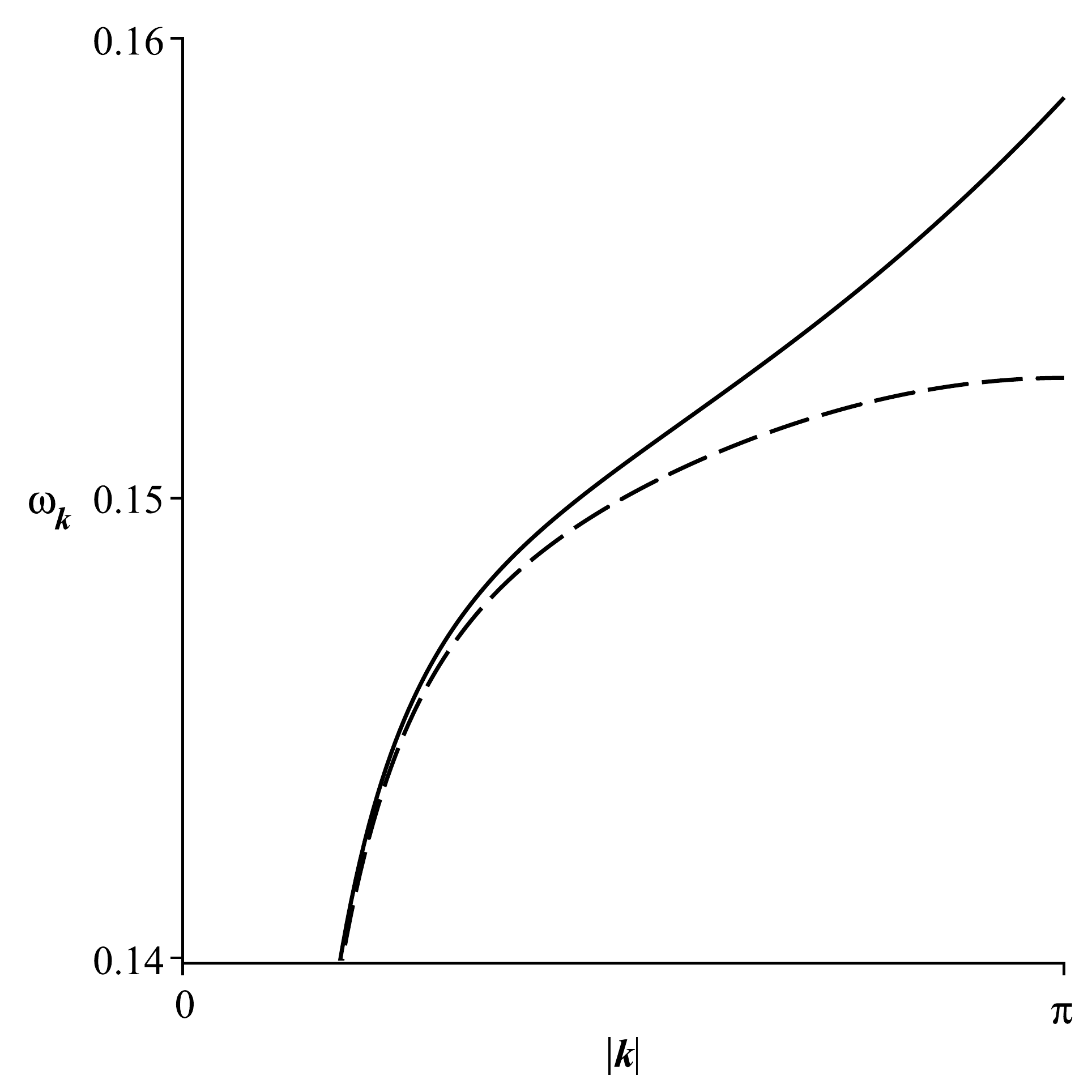}} \newline
\caption{\textcolor{black}{\textbf{Dispersion relations for $R = 0.15$.}
Comparison between the
approximated (solid line) and exact (dashed line) solutions for the acoustic branch (a zoom at the end
of the Brillouin zone, in order to reveal their slight discrepancy):  (a) disordered atoms (graphic representation of eq. \eqref{omega_D} vs. \eqref{omega_D_app}); (b) ordered atoms (graphic representation of eq. \eqref{omega_O} vs. \eqref{omega_O_app}).
}}
\label{completapprox}
\end{figure}

It is important to emphasize that the expressions obtained in the small-angle approximation (i.e. $\sin(ak_i)\approx ak_i$) are valid with excellent accuracy over the whole Brillouin zone, due to the smallness of the $R$-parameter (for $R=0.1$, the ratio between the approximate and the exact values of the frequency at the border of the Brillouin zone, where the discrepancy is maximal, is about 1.5 \%), which can be clearly seen in the panels a and b of Figure~\ref{completapprox}.

Both functions \eqref{omega_D} and \eqref{omega_D_app} are monotonically
increasing within the Brillouin zone, with the maximum at $ak_z=\pi$. In the panels c and d of Figure~\ref{completapprox} we depicted the dispersion relations for $R=0.15$.
We observe the characteristic plateau in $\omega _{\mathbf{k}}$, which leads to
the broadened van Hove singularity in the VDOS.
The small angle approximation gives a very good estimate of the full solution for small $R$,
therefore we work further with the approximate dispersion relation 
\eqref{omega_D_app}.

We may cast equation \eqref{disordered_se} in the form
\begin{equation}
k_{x}^{2}+k_{y}^{2}+k_{z}^{2}=\frac{1}{a^{2}}\frac{3\omega_{\mathbf{k}}
^{2}R^{2}-\frac{\omega_{\mathbf{k}} ^{4}}{\omega _{0}^{2}}}{\omega
_{0}^{2}R^{2}-\omega_{\mathbf{k}} ^{2}\left( 1+R^{2}\right) }
\label{volume_disordered}
\end{equation}%
and identify the isofrequency surfaces $\omega_{\mathbf{k}}=\omega=\text {const.}$ in the ${\mathbf{k}}$-space as spheres of radius $\frac{1}{a^{2}}\frac{3\omega ^{2}R^{2}-\frac{\omega ^{4}}{\omega _{0}^{2}}}{\omega _{0}^{2}R^{2}-\omega ^{2}\left( 1+R^{2}\right) }$.
The density of states is obtained by the differentiation of the volume of the sphere with respect to $\omega$ and subsequently dividing it by the volume of one cell in the ${%
\mathbf{k}}$-space, $(2\pi)^3/V$, where $V$ is the volume of the system in
the direct space. Thus the density of states for the disordered atoms reads
\begin{align}
g_{\rm dis}\left( \omega \right) =& \frac{3}{2} \frac{V}{ \left( 2\pi
\right) ^{3}}\frac{4\pi \omega ^{2}}{\omega _{0}^{3}a^{3}}\sqrt{\left\vert
\frac{3-\frac{\omega ^{2}}{\omega _{0}^{2}R^{2}}}{1-\frac{\omega ^{2}}{%
\omega _{0}^{2}R^{2}}}\right\vert }
 \left( \frac{\omega ^{2}}{\omega
_{0}^{2}R^{2}}\right) 
\left\vert \frac{\frac{\omega ^{2}}{\omega
_{0}^{2}R^2}\left(1-\frac{3R^2}{2}\right)-2}{\left( 1-%
\frac{\omega ^{2}}{\omega _{0}^{2}R^{2}}\right) ^{2}}\right\vert ,
\label{38}
\end{align}
where the prefactor 3/2 accounts for the three acoustic branches and the fact that only half of the atoms are disordered.

\subsection{Density of states for ordered atoms}

In a similar fashion as described in the preceding subsection, we start the analysis for the ordered atoms. We  emphasize that the distinction ordered/disordered atoms is a matter of semantics, because all atoms suffer the
effects of the noncommutativity of coordinates either directly or through the dynamical couplings.
If the atom $(lmn)$ is a ordered one, its coordinates are $\widehat u^i_{lmn}=\widehat U^i_{lmn}$, while its nearest neighbors are disordered and the shift of noncommutative coordinates has to be applied to them, namely:
\begin{align}
\widehat{u}^x_{l\pm1mn}&=\widehat U^x_{l\pm1mn}-\frac{1}{2\hbar}%
\theta_{xj}\widehat P_{l\pm1mn}^{j}, \ \ j=y,z,\cr 
\widehat{u} ^y_{lm\pm1n}&=\widehat U^y_{lm\pm1n}-\frac{1}{2\hbar}\theta_{yj}\widehat
P_{lm\pm1n}^{j}, \ \ j=z,x,\cr
\widehat{u}^z_{lmn\pm1}&=\widehat
U^z_{lmn\pm1}-\frac{1}{2\hbar}\theta_{zj}\widehat P_{lmn\pm1}^{j}, \ \
j=x,y.
\end{align}
Proceeding exactly as in the case of disordered atoms, we find the equations of
motion for ordered atoms up to the second order in $R$:
\begin{align}  \label{O_solution}
\ddot{\widehat U}^x_{lmn}  =&-\omega_{0}^{2}\left(2{\widehat U}^x_{lmn}-{%
\widehat U}^x_{l-1mn}-{\widehat U}^x_{l+1mn}\right)+\frac{\omega_{0}R}{2}%
\biggl\{\dot{\widehat U}^z_{l-1mn}+\dot{\widehat U}^z_{l+1mn}-\dot{\widehat U%
}^y_{l-1mn}-\dot{\widehat U}^y_{l+1mn}\biggr\}\cr
& +\frac{\omega_{0}^{2}R^{2}}{4}\biggl\{-2\left[2{\widehat U}^x_{l-1mn}-{\widehat U}%
^x_{l-2mn}-{\widehat U}^x_{lmn}\right]-2\left[2{\widehat U}^x_{l+1mn}-{%
\widehat U}^x_{l+2mn}-{\widehat U}^x_{lmn}\right]\cr
 & +\left[2{\widehat U} ^y_{l-1mn}-{\widehat U}^y_{l-1m-1n}-{\widehat U}^y_{l-1m+1n}\right]+\left[2{\widehat U}^y_{l+1mn}-{\widehat U}^y_{l+1m-1n}-{\widehat U}^y_{l+1m+1n}%
\right]\cr
 & + \left[2{\widehat U}^z_{l-1mn}-{\widehat U}^z_{l-1mn-1}-{%
\widehat U}^z_{l-1mn+1}\right]+\left[2{\widehat U}^z_{l+1mn}-{\widehat U}%
^z_{l+1m+2n-1}-{\widehat U}^z_{l+1mn+1}\right]\biggr\}.
\end{align}
The remaining equations of motion for the components $(y,z)$ can readily be obtained, but since they have the same structure as \eqref{O_solution} and are rather lengthy we shall omit them. 

We use again the Fourier expansion \eqref{Fourier_solution} and find the the elements of the matrix of coefficients $A_{ij}$, explicitly given by the formulas \eqref{saec_eq_ordered}.
The saecular equation for the ordered atoms is given by
\begin{equation} \label{ord_se}
\det|A_{ij}|=0.
\end{equation}
Upon isotropization, the exact solutions of \eqref{ord_se} are:
\begin{align}
\omega _{{\mathbf{k}}\pm }^{\mathrm{O}} =&\frac{\omega _{0}}{2}\bigg[
6R^{2}+8\sin ^{2}\left( \frac{a|{\mathbf{k}}|}{2}\right) \left[ 1-2R^{2}\sin
^{2}\left( \frac{a|{\mathbf{k}}|}{2}\right) \right]     \cr
&  \pm 2\bigg[ 64R^{2}\sin ^{6}\left( \frac{a|{\mathbf{k}}|}{2}\right) %
\left[ R^{2}\sin ^{2}\left( \frac{a|{\mathbf{k}}|}{2}\right) -1\right] \cr
&+\left( 16-48R^{4}\right) \sin ^{4}\left( \frac{a|{\mathbf{k}}|}{2}\right)
+8R^{2}\sin ^{2}\left( \frac{a|{\mathbf{k}}|}{2}\right) +9R^{4}\bigg] ^{%
\frac{1}{2}}\bigg] ^{\frac{1}{2}}.  \label{omega_O}
\end{align}

In the small angle approximation, the saecular equation for ordered atoms becomes
\begin{equation}
\omega_{\mathbf{k}} ^{4}-\omega _{0}^{2}a^{2}\left( \omega
_{0}^{2}R^{2}-\omega_{\mathbf{k}} ^{2}\right) |\mathbf{k}|^2 +3\omega_{%
\mathbf{k}} ^{2}\omega _{0}^{2}R^{2}=0,  \label{ordered_se}
\end{equation}
or equivalently
\begin{equation}
k_{x}^{2}+k_{y}^{2}+k_{z}^{2}=\frac{1}{a^{2}}\frac{3\omega ^{2}R^{2}-\frac{%
\omega ^{4}}{\omega _{0}^{2}}}{\omega _{0}^{2}R^{2}-\omega ^{2}}.
\label{volume_ordered}
\end{equation}
The solutions of the saecular equation  \eqref{ordered_se} are:
\begin{align}
\omega _{{\mathbf{k}}\pm }^{\mathrm{O-approx}} =&\frac{\omega _{0}}{2}\bigg[
6R^{2}+2a^{2}|{\mathbf{k}}|^{2}\left( 1-\frac{R^{2}}{2}a^{2}|{\mathbf{k}}%
|^{2}\right) \cr
& \pm 2\sqrt{R^{2}a^{6}|{\mathbf{k}}|^{6}\left( \frac{R^{2}}{4}a^{2}|{%
\mathbf{k}}|^{2}-1\right) +\left( 1-3R^{4}\right) a^{4}|{\mathbf{k}}%
|^{4}+2R^{2}a^{2}|{\mathbf{k}}|^{2}+9R^{4}}\bigg] ^{\frac{1}{2}}. \label{omega_O_app}
\end{align}
Using the same procedure as for the disordered atoms, the contribution of the ordered atoms to the density of states is found to be:
\begin{equation}
g_{\rm ord}\left( \omega \right) =\frac{3}{2}\frac{V}{(2\pi)^2
}{\frac{1}{a^3}}\frac{\omega ^{2}}{\omega _{0}^{3}}\sqrt{\left\vert \frac{3-\frac{\omega
^{2}}{\omega _{0}^{2}R^{2}}}{1-\frac{\omega ^{2}}{\omega _{0}^{2}R^{2}}}%
\right\vert }. \label{g_order}
\end{equation}

Now that we have determined the full isotropised density of states, given by the sum of the density of states of the two kinds of atoms present in the lattice, we proceed next to the discussion of the boson peak.


\section{Reduced specific heat and boson peak}
\label{sec.4}


The complete glass density of states in the small angle approximation is equal to the sum of the  disordered and ordered atoms expressions, Eqs.~\eqref{38} and \eqref{g_order}, respectively, namely
\begin{equation}\label{g_glass_1}
g_{\rm glass}\left( \omega \right) =\frac{3}{2}\frac{V}{(2\pi)^2
}{\frac{1}{a^3}}\frac{\omega ^{2}}{\omega _{0}^{3}}\sqrt{\left\vert \frac{3-\frac{\omega
^{2}}{\omega _{0}^{2}R^{2}}}{1-\frac{\omega ^{2}}{\omega _{0}^{2}R^{2}}}%
\right\vert }
 \left[ 1+\frac{%
\omega ^{2}}{\omega _{0}^{2}R^{2}}\left\vert \frac{\frac{\omega ^{2}}{\omega
_{0}^{2}R^{2}}-2-\frac{3}{2}\frac{\omega ^{2}}{\omega _{0}^{2}}}{\left( 1-%
\frac{\omega ^{2}}{\omega _{0}^{2}R^{2}}\right) ^{2}}\right\vert \right].
\end{equation}%
It is important to emphasize the presence of a divergence in the VDOS, i.e. a van Hove singularity, which occurs for $\omega _{\mathrm{div}}=\omega _{0}R$, that is purely an effect of the fuzziness and isotropy of this glass model.
In the limit $\theta \rightarrow 0$ (or equivalently $R \to 0$), we recover the VDOS of a usual simple cubic lattice with one atom per cell.

\begin{figure}[htp!]
\centering \includegraphics[width=10cm]{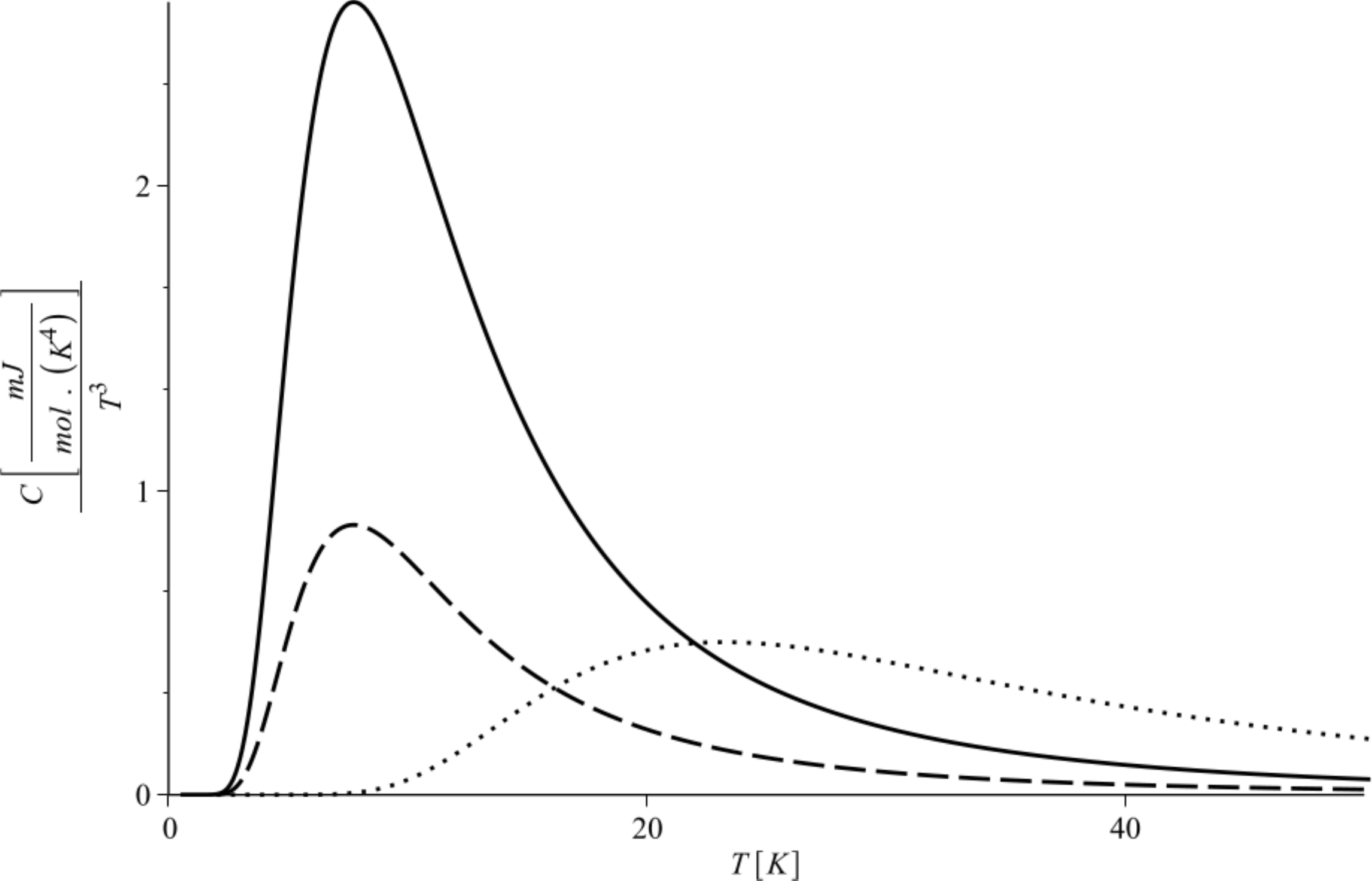} 
\caption{\textcolor{black}{Different theoretical predictions for the reduced specific heat curve versus temperature around the boson peak as well as the behaviour of the main parameters of the model. Parameters: $R=0.1,~\omega_{div}=5$THz for the solid curve, $R=0.07,~\omega_{div}=5$THz for the dashed curve and $R=0.1,~\omega_{div}=10$THz for the dotted curve.
}}
\label{FigureB}
\end{figure}

The reduced specific heat is determined from the following expression (see, e.g., \cite{Kantorovich})
\begin{equation}\label{specific_heat_1}
\frac{C}{T^3}=\int_0^{\omega_{\rm{max}}} \frac{ \hbar^2 \omega^2}{k_{B}T^5}\frac{N_A}{Z} \frac{e^{\hbar\omega/k_{B}T}}{(e^{\hbar\omega/k_{B}T}-1)^2} g_{\rm glass}(\omega) d\omega,
\end{equation}
in which $Z$ is the number of formula units per unit cell (in our model, $Z=1$) and $k_B$ is the Boltzmann constant.
We take $\omega _{\mathrm{max} }=\sqrt 3\omega _{0}R$ in \eqref{specific_heat_1}, that is the frequency of the optical branches at $\mathbf{k}=\vec{0}$; in other words, we have included only the acoustic modes and left aside the optical ones, for which the small angle approximation works only for an insignificant number of modes.
The boson peak frequency is clearly present in the acoustic branches (panels c and d in Figure~\ref{completapprox}), therefore this omission does not affect our region of interest.
The universality of the model in describing the boson peak phenomena for different types of glasses can be observed in Figure~\ref{FigureB}.
On the one hand, we can see in Fig.~\ref{FigureB} that the parameter $R$ is responsible for the peak's amplitude (see for comparison the solid and dashed curves); on the other hand, the parameter $\omega_{\rm div}$ accounts for the peak's position in terms of temperature.

\begin{figure*}[ht]
\subfloat[Subfigure 1 list of figures text][Amorphous germanium dioxide]  {\
\includegraphics[width=8cm]{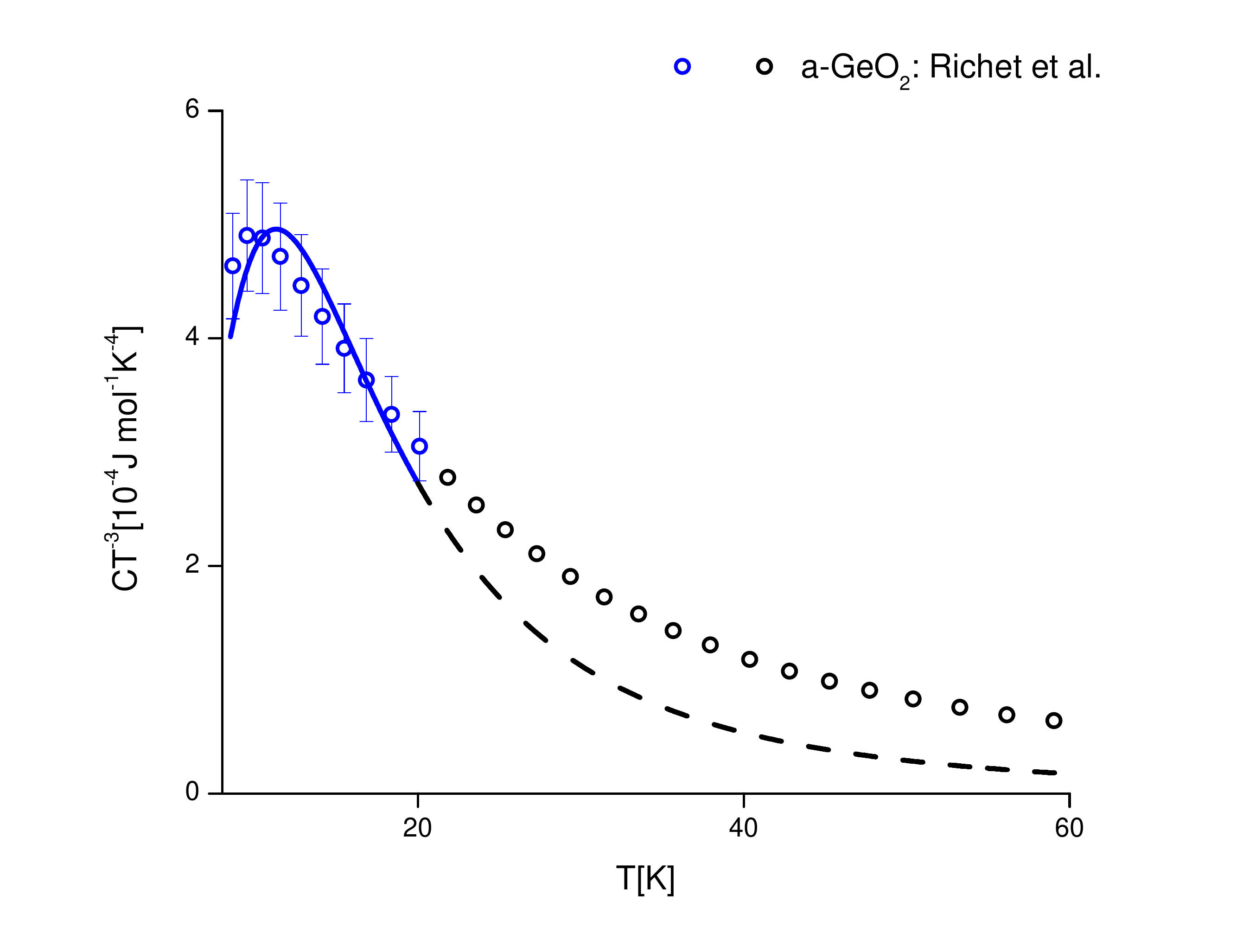}} ~
\subfloat[Subfigure 1 list of figures text][$\alpha -n-$Ba$_{8}$Ga$_{16}$Sn$_{30}$ clathrate]  {\
\includegraphics[width=8cm]{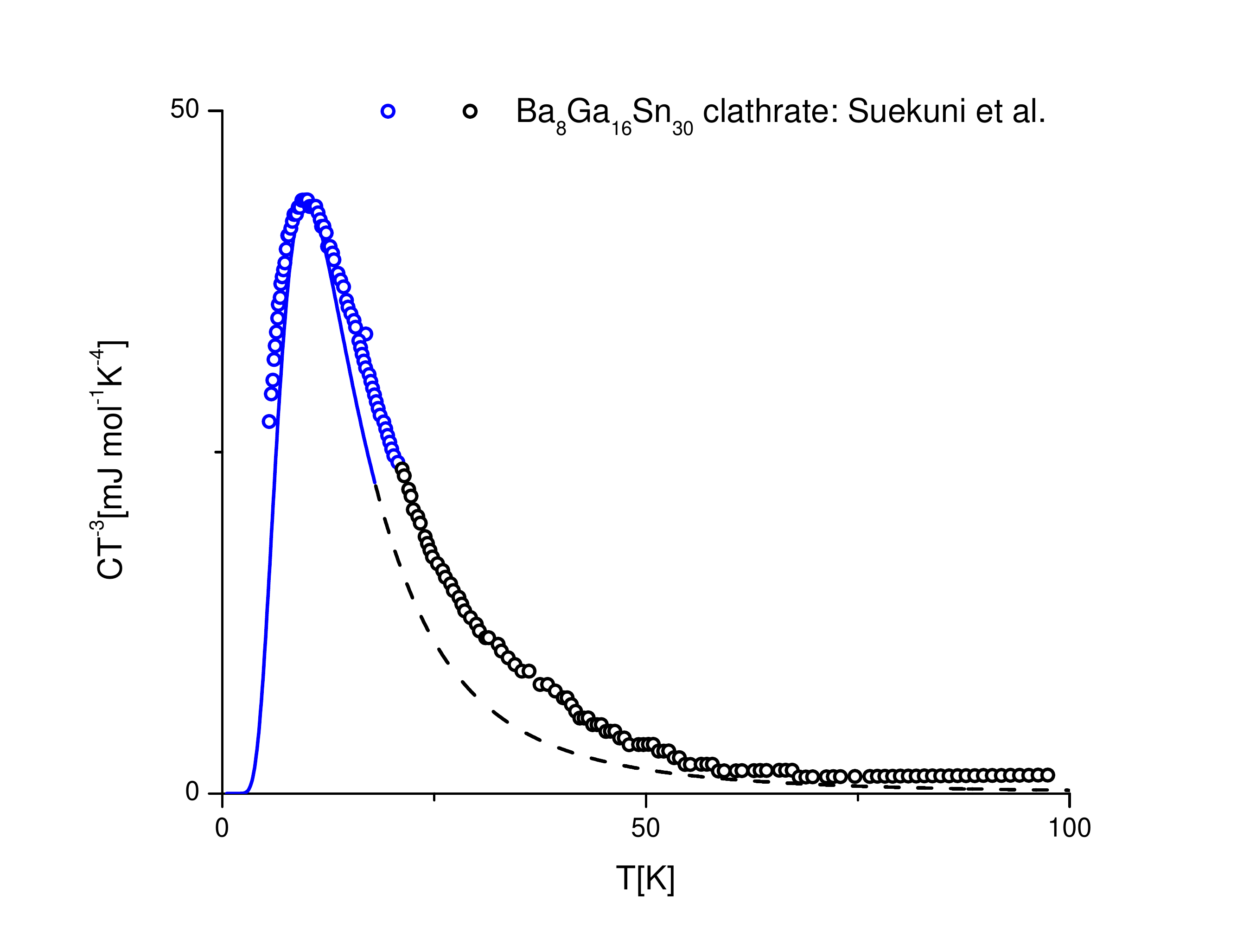}}
\caption{\textbf{Experimental data versus theoretical prediction.} Experimental data represented as empty balls for (a) amorphous germanium dioxide ($a$-GeO$_{2}$), with $10 \%$ standard deviation, as reported in  \cite{GeO2}; (b) $\alpha-n-$Ba$_{8}$Ga$_{16}$Sn$_{30}$ clathrate \cite{BGS}. The theoretical prediction according to \eqref{specific_heat_1} is represented by the solid blue line up to $\sqrt 3 T_{\mathrm{peak}} $ and black dashed line at higher temperatures.} \label{FigureA}
\end{figure*}

Noncommutativity is the key feature in our description of amorphous solids, because it is responsible for engendering a divergence in the density of states in the acoustic branches, i.e. a van Hove singularity. 
It is known that when this divergence is present in the expression for the specific heat \eqref{specific_heat_1} we observe a departure from the ordinary Debye's theory for crystalline solids, which corresponds exactly to the expected profile of glasses: excess of heat capacity and the boson peak phenomena (see Figure~\ref{FigureA}).

To validate our small-$R$ approximation, we have applied the model for the case of the amorphous germanium dioxide ($a$-GeO$_{2}$)  \cite{GeO2}. The model has two free parameters: the characteristic frequency  $\omega_0$ and the noncommutativity parameter $\theta$.  They are determined by fitting the theoretical curve \eqref{specific_heat_1} to the experimental data, such that the frequency and reduced specific heat at the peak match. 

In figure~\ref{FigureA} is shown the agreement around the peak between the experimental curves and the theoretical predictions of our model based on liquid-type disorder effects. The small-angle approximation and the exclusion of the optical modes limit the theoretical prediction according to formula \eqref{specific_heat_1} to temperatures of up to~$\sqrt 3 T_{\mathrm{peak}}$. This is indicated in figure~\ref{FigureA} by the blue color. The black color marks the region where our present estimation of the reduced specific heat is imprecise due to the exclusion of the optical modes in the integral, among other simplifying assumptions. Nevertheless, they exist in the model and can be included by numerical methods. The regime of temperatures below $T_{\mathrm{peak}}$ is characterized by the appearance of tunneling two-level systems \cite{Phillips, Anderson} and/or other possible phenomena \cite{Leggett}, which also are not included in \eqref{specific_heat_1}. Although we have no data for this glass much below $T_{\mathrm{peak}}$, we would still expect a departure of the theoretical curve from the possible experimental data in that region.

\begin{table*}[htb]
\resizebox{\textwidth}{!}{%
\begin{tabular}{|c|c|c|c|c|c|c|c|c|}
\hline
Glassy material & $R$ & $\omega _{0}\left( \times 10^{14}\right) $ & $\omega
_{\theta }\left( \times 10^{15}\right) $ & $\omega _{\mathrm{div}}\left(
\times 10^{12}\right) $ & $2\pi \theta \left( \times 10^{-24}\right) $ & $%
N\left( \times 10^{28}\right) $ & $a^{2}\left( \times 10^{-20}\right) $ & $%
T_{\mathrm{peak}}$ \\  \hline
$a$-GeO$_{2}$ & $0.0697$ & $1.047$ & $1.503$ & $7.300$ & $2.539$ & $2.655$ &
$8.066$ & $11.2$ \\ \hline
$\alpha n$BGS clathrate & $0.290$ & $0.241$ & $0.083$ & $7.000$ & $31.504$ &
$0.392$ & $34.106$ & $10.32$ \\ \hline
\end{tabular}}
\caption{Characteristic parameters for  a-GeO$_{2}$ and $\alpha-n-$Ba$_{8}$Ga$_{16}$Sn$_{30}$ clathrate. All frequencies are in rad s$^{-1}$. The value of $\theta $ (in m$^{-2}$) is determined from the relation $\omega _{\mathrm{div} }=\omega _{0}R$, using the experimental curves. $N$ is the number density in m$^{-3}$, whereas $a$ is the lattice spacing for our model in m$^{2}$. $T_{\mathrm{peak}}$ is the temperature at the peak of the reduced specific heat in K.}
\label{TableA}
\end{table*}

\section{Discussion and outlook}
\label{sec.5}

Glasses present anomalous low-temperature characteristics, generically known as the "boson peak", whose description in a first-principle theoretical framework has been under debate over many years. So far, the structure and dynamics
of glasses has been described in the literature on the basis of phenomenological models. The interpretations of experimental studies have led to numerous hypotheses:  the glass anomaly could be assigned to structural motifs with (quasi-)localized transverse vibrational
modes that resonantly couple with transverse phonons and thus
govern their dissipation \cite{shintani}, or to a broadening of the transverse acoustic van Hove singularity of the corresponding crystal \cite{chumakov}, or to the presence of additional vibrational modes which can be induced solely by structural disorder  \cite{Brink}. 
However, none of the known phenomenological approaches to this problem has allowed the derivation of a complete, widely-accepted theory of amorphous solids \cite{27,elliott,duval,klinger,buchenau,tanaka,grigera,gotze,
lubchenko,silbert}. 

In this paper, we have proposed to interpret an amorphous solid as a system with a frozen-in liquid-type of disorder, implemented mathematically as a noncommutative algebra of coordinate operators. Intuitively, this is equivalent to a fuzziness or uncertainty in the positions of the particles that form the glass. However, this is not a simple positional disorder, as the delocalization depends essentially on the momenta of the particles -- a feature specific to noncommuting coordinates. The quantum mechanical model we propose is developed from first principles and permits the derivation of {\it analytic} formulas for the density of states and specific heat of glasses. Other important features of the model are its simplicity and the very small number of free parameters ($\omega_0$ and $\theta$, the former connected to the electromagnetic interactions among the glass particles and the latter being a measure of the disorder). This novel theory naturally accounts for the excess of heat capacity and the boson peak phenomena, which are manifestations of a pronounced divergence in the density of states in the acoustic branches, i.e. a van Hove singularity.

In the Debye theory of the specific heat of crystals, the contributing modes at sufficiently low temperature are long wavelength acoustic ones, whose excitation is approximately classical. Nevertheless, the quantum mechanical principles apply, according to which the excited modes follow a Bose--Einstein distribution. As far as the spectrum of the low-temperature excitations, one can in principle use either a classical treatment or a  quantum mechanical (phonon) treatment, with the same reasults \cite{Landau}. In our model of glasses, the same long wavelength acoustic modes contribute essentially to the specific heat in the region of the boson peak. Technically, due to the quantization of space encoded in the noncommuting coordinates, the treatment is quantum mechanical throughout. However, we emphasize that it is not the usual quantum mechanics which makes the difference in the calculation of the spectrum, but the additional quantization of coordinates that models the disorder. In effect, the two constants $\hbar$ and $\theta$ appear always together in a factor $R$ (see \eqref{R}), and in the limit $\theta\to0$ one recovers the usual crystalline spectrum, with no reference to the Planck constant. 

Glasses are complex systems, and their thermal behaviour at different temperatures is dominated by different aspects of their structure and interactions. In the present formulation, our model gives reliable results in the same range as the Debye theory for crystals, i.e. for temperatures below $0.1 T_D$ (see \cite{Kittel}), where $T_D$ is the Debye temperature of the crystal. It is well known that the Debye model is accurate in the low-temperature and high-temperature regimes, but not in the range of intermediate temperatures, due to its simplifying assumptions. Our model has similar limitations, but it covers well the region of the boson peak, where Debye's $T^3$-law applies in the case of crystals. 

Our present model of the ideal glass spectrum of vibrations at low temperatures is an entirely analytical glass model, not requiring computer simulations. It provides a robust framework for the inclusion of further complexities. The model can be refined by including the next-to-nearest neighbor couplings, in which case we expect to see the peak in the transverse acoustic branch. It is also interesting to study whether adopting in the model the actual lattice type of the crystal counterpart of the analyzed glasses would improve the agreement with experimental data. Further confirmations have to include the calculation of theoretical predictions and the comparison with data for other quantities whose the anomalous behaviour is assigned to the boson peak, like, for example, the thermal conductivity and the Raman spectra in the THz region. Incidentally, crystals with defects, which also present the boson peak (see, e.g., \cite{Parshin, Milkus}), can also be modeled using this approach, by increasing appropriately the number of ordered motifs in between two disordered ones, depending on the degree of amorphization.
A more detailed analysis of the structural and dynamical aspects of the model is in progress.

\section*{Acknowledgments}
The authors are grateful to M. Chaichian for valuable comments. T.~Cardoso e Bufalo thanks B.~M.~Pimentel and H.~S.~Martinho for discussions.
The financial support of FAPESP through Project no. $2016/03921-8$, and the Academy of Finland through the Projects no. 136539 and 272919 is acknowledged.
R.B. acknowledges partial support from Conselho
Nacional de Desenvolvimento Cient\'ifico e Tecnol\'ogico (CNPq Projects No. 305427/2019-9 and No. 421886/2018-8) and Funda\c{c}\~ao de
Amparo \`a Pesquisa do Estado de Minas Gerais (FAPEMIG Project No. APQ-01142-17).
\appendix

\section{Elements of the matrix of coefficients $A_{ij}$ for ordered atoms}

The explicit expressions for the elements of the matrix of coefficients $A_{ij}$  \eqref{ord_se} for the ordered atoms:
\begin{align}
A_{11}&=\omega_{\mathbf{k}} ^{2}-4\omega _{0}^{2}\sin ^{2}\left( \frac{%
ak_{x}}{2}\right) +2\omega _{0}^{2}R^{2}   \Psi (k_x)  , \cr
A_{12}&= \left[-i\omega _{0}\omega_{\mathbf{k}} R  +2\omega _{0}^{2}R^{2} \sin ^{2}\left(
\frac{ak_{y}}{2}\right) \right] \Phi (k_x), \cr
A_{13}&= \left[i\omega _{0}\omega_{\mathbf{k}} R   +2\omega _{0}^{2}R^{2} \sin ^{2}\left(
\frac{ak_{z}}{2}\right) \right] \Phi (k_x), \cr
A_{21} &= \left[i\omega_{\mathbf{k}} \omega _{0}R   +2\omega _{0}^{2}R^{2}\sin ^{2}\left(
\frac{ak_{x}}{2}\right) \right]\Phi (k_y),\cr
A_{22} &=\omega_{\mathbf{k}} ^{2}-4\omega _{0}^{2}\sin ^{2}\left( \frac{%
ak_{y}}{2}\right) +2\omega _{0}^{2}R^{2}  \Psi (k_y),\cr
A_{23} &= \left[-i\omega_{\mathbf{k}} \omega _{0}R   +2\omega _{0}^{2}R^{2} \sin ^{2}\left(
\frac{ak_{z}}{2}\right) \right] \Phi (k_y) , \cr
A_{31} &=\left[-i\omega_{\mathbf{k}} \omega _{0}R  +2\omega _{0}^{2}R^{2} \sin ^{2}\left( \frac{%
ak_{x}}{2}\right) \right] \Phi (k_z) , \cr 
A_{32} &=\left[ i\omega_{\mathbf{k}} \omega _{0}R   +2\omega _{0}^{2}R^{2} \sin ^{2}\left( \frac{%
ak_{y}}{2}\right) \right] \Phi (k_z) ,  \cr
A_{33} &=\omega_{\mathbf{k}} ^{2}-4\omega _{0}^{2}\sin ^{2}\left( \frac{%
ak_{z}}{2}\right) +2\omega _{0}^{2}R^{2}  \Psi (k_z).  \label{saec_eq_ordered}  
\end{align}
where we have defined, by simplicity of notation, the following symbols
\begin{align}
\Psi (k_i) & =  2\sin ^{2}\left( \frac{ak_{i} }{2}\right) 
-\sin ^{2}\left( ak_{i}\right),   \cr
\Phi (k_i) & = 1 -2  \sin ^{2}\left( \frac{ak_{i}}{2}\right).
\end{align}

\end{document}